\documentclass[reprint,twocolumn,aps,prc,a4paper,superscriptaddress,showpacs,preprintnumbers,amsmath,amssymb]{revtex4}
\usepackage{CJK} 
\usepackage{color}
\usepackage[colorlinks,linkcolor=blue,anchorcolor=blue,citecolor=blue]{hyperref}
\usepackage{graphicx}% Include figure files
\usepackage{dcolumn}% Align table columns on decimal point
\usepackage{bm}% bold math
\begin{document}
%\bibliographystyle{apsrev4-1}
%\begin{CJK}{GBK}{kai}  
%\preprint{APS/123-QED}

\title{Correlating charge radius with quadrupole deformation and $B(E2)$ in atomic nuclei}

%\author{Ann Author}
% \altaffiliation[Also at ]{Physics Department, XYZ University.}%Lines break automatically or can be forced with \\
%\author{Second Author}%
% \email{Second.Author@institution.edu}
%\affiliation{%
% Authors' institution and/or address\\
% This line break forced with \textbackslash\textbackslash
%}%

%\collaboration{MUSO Collaboration}%\noaffiliation
\author{Bao-Hua Sun}\thanks{Corresponding author: bhsun@buaa.edu.cn}
\affiliation{School of Physics and Nuclear Energy Engineering, Beihang University, Beijing 100191, China}
\affiliation{International Research Center for Nuclei and Particles in the Cosmos, Beijing 100191, China}
\author{Chuan-Ye Liu} 
\affiliation{School of Physics and Nuclear Energy Engineering, Beihang University, Beijing 100191, China}
\author{Hao-Xin Wang}
\affiliation{School of Physics and Nuclear Energy Engineering, Beihang University, Beijing 100191, China}

%\collaboration{CLEO Collaboration}%\noaffiliation
 
\date{\today}% It is always \today, today,
             %  but any date may be explicitly specified

\begin{abstract}
A good linear correlation is found between the four-point charge radius relation $\delta R_{2p-2n}(Z,N)$ with 
that of quadrupole deformation data in even-even nuclei. 
This results in a further improved charge radius relation that holds in a precision of about  5$\times 10^{-3}$ fm. 
The new relation can be generalized to the reduced electric quadrupole transition probability $B(E2)$ 
between the first $2^+$ state and the $0^+$ ground state, and the mean lifetime $\tau$ of the first 2$^+$ state. 
Same correlations are also seen in global nuclear models, their precisions, however, are not  enough to be consistent with the experimental data. 

%\begin{description}
%\item[Usage]
%Secondary publications and information retrieval purposes.
%\item[PACS numbers]
%May be entered using the \verb+\pacs{#1}+ command.
%\item[Structure]
%You may use the \texttt{description} environment to structure your abstract;
%use the optional argument of the \verb+\item+ command to give the category of each item.
%\end{description}
\end{abstract}
\pacs{21.10.Ft, 21.10.Tg, 23.20.-g, 29.87.+g}
%\pacs{Valid PACS appear here}% PACS, the Physics and Astronomy
                             % Classification Scheme.
%\keywords{Suggested keywords}%Use showkeys class option if keyword
                              %display desired
\maketitle

%\tableofcontents

\section{Introduction}

Mass, charge radius, lifetime, electric (magnetic) transition probability, and deformation are among the most fundamental observables for the many-body nuclear system. 
A systematic analysis of these data have been successful in bringing forth a global picture of the atomic nuclei. 
For example, experimental binding energies or charge radii for neighboring nuclei do not differ much 
except at several specified regions like closed shells or onset of shape. 

On the other hand, comparing these observables of atomic nuclei, which differ by one or a few neutrons or protons, 
have yielded many empirical relations or filters for special interaction strengths between the valence nucleons. 
Of them, the Garvey-Kelson relations for nuclear binding energies are probably one of the best known 
examples~\cite{GK1966PRL16,GK-np2001,GK2008PhysRevC.77.041304,GK2011PhysRevC.83.054309,GK2013-PhysRevC.87.057304,GK2013PhysRevC.87.014313,GK2013PhysRevC.88.064325,GK2014PhysRevC.89.061304}. 
The validity of this relation in the nuclear charge radius has  also been examined recently~\cite{Piekar2010EPJA46}.

The radius of the charge (proton) distribution can be assumed to be equal to that  of the nuclear mass distribution, 
considering the nucleus as a liquid drop with the protons homogeneously distributed over the sphere of the nucleus.  
Although not accurate in prediction, this simple liquid-drop model can serve as a guide, and  the interesting physics
can be found in local deviations from the global behavior.
In our recent work~\cite{Sun2014PhysRevC.90.054318,Sun2015PhysRevC.91.019902} 
we proposed a set of nuclear charge radius relations $\delta R_{ip-jn}(Z,N)$, 
\begin{eqnarray}
   \delta R_{ip-jn}(Z,N)&= &R(Z,N)-R(Z,N-j)     \nonumber \\
                        & &-[R(Z-i,N)-R(Z-i,N-j)]    \nonumber \\
                        &\simeq & 0  \;,
   \label{eq1}
\end{eqnarray}
where $R(Z,N)$ is the root-mean-square (rms) charge radius of the nucleus with $N$ neutrons and $Z$ protons. $i$ and $j$ are integers.  
The validity of such relation  is a consequence of the smooth transition in the nuclear structure that is often found when
going from a nucleus to its neighboring nuclei.  Eq.~\ref{eq1} holds precisely over almost the whole nuclear chart
except at a few regions characterized by shape transition and shape coexistence at, e.g., $N \sim 60$, $N \sim 90$ and $Z \sim 80$. 
These exceptions raise the possibilities that more accurate local systematics may be developed from the experimental data. 

One simple case connecting only even-even nuclei is 
\begin{eqnarray}
   \delta R_{2p-2n}(Z,N)&= &R(Z,N)-R(Z,N-2)    \nonumber \\
                        & &-[R(Z-2,N)-R(Z-2,N-2)]   \nonumber \\ 
                        &\simeq & 0  \;.
   \label{eq2}
\end{eqnarray}
The term $R(Z,N)-R(Z,N-2)$, the so-called isotope shift, 
involves the variation of the charge distribution when only two neutrons are added to the system. 
In this sense, $\delta R_{2p-2n}(Z,N)$ is nothing but 
the difference of isotope shifts for neighboring two isotopic chains. Hereafter we simply 
rewrite $\delta R_{2p-2n}(Z,N)$ as $\delta R(Z,N)$.

In this work, we aim to examine and quantify the correlation between the local charge radius relations  $\delta R(Z,N)$
and those of deformation data.  The correlation is made from cases that both charge radius and quadrupole deformation data are experimentally availbale. 
This then leads us to an improved relation by correcting the contribution from quadrupole deformation effect in atomic nuclei. 
Moreover, this new relation can be naturally extended to the reduced electric quadrupole transition probability $B(E2)$ between the first $2^+$ state and the $0^+$ ground state, 
and the mean lifetime $\tau$ of the first 2$^+$ state.

\section{Shape effect on charge radii}\label{sec1}

For the system of spherical nuclei, the rms charge radii can be empirically described by 
\begin{equation}\label{eq:Alaw}
R(Z,N)=\sqrt{3/5} r_0 A^{1/3} \;, 
\end{equation}
where $A$ is the mass number and $r_0$ is fixed to 1.2 fm throughout this paper.   
Thus $\delta R(Z,N)$ for the even-even isotopes is
\begin{equation}
   \delta R(Z,N)   \equiv  \delta R(Z,N)_{\mbox{sph}} = \sqrt{\frac{3}{5}}r_0\delta(A^{1/3}) \;. 
   \label{eq4}
\end{equation}
Numerically it is easy to see that  $\delta R(Z,N)_{\mbox{sph}}$ goes down to a few times $10^{-4}$ fm with increasing mass number. 

For a deformed nucleus, the charge  radius can be simply decoupled to the spherical and deformation part.   
For the important case of an axially symmetric shape,  
by neglecting the high-order corrections one can express this approximately as: 
\begin{equation}\label{eq.5}
 R(Z,N)=R_0(Z,N)\left[1+\frac{5}{8\pi}\beta_2^2(Z,N)\right] \; 
\end{equation}
where $\beta_2(Z,N)$ is the rms quadrupole deformation for the nuclide ($Z,N$). As will be described later,  it is derived experimentally from the  reduced electric quadrupole transition probability $B(E2)$. 
 $R_0(Z,N)$ corresponds to the charge radius of a spherical nucleus that the nuclear volume is conserved, and is defined by Eq.~\ref{eq:Alaw}.    
Eq.~(\ref{eq.5}) may be generalized to include higher order multipoles or triaxial shape~\cite{Bohr1969book,Greiner1996book}. 
 
Accordingly, the experimental four-point relation $\delta R(Z,N)$ can be expressed in terms of two variables representing the spherical equivalent radius and the deformation.
This corresponds to a two-parameter model of describing $\delta R(Z,N)$, 
\begin{eqnarray}
   \delta R(Z,N)   & = &  \delta R(Z,N)_{\mbox{sph}}   +  \delta R(Z,N)_{\mbox{def}}  \nonumber \\
   & = & \delta R(Z,N)_{\mbox{sph}} + \frac{5}{8\pi}\delta(R_0\beta_2^2)  \;,
   \label{eq6}
\end{eqnarray}
where $ \delta R(Z,N)_{\mbox{sph}}$ is defined in Eq.~(\ref{eq4}), and $\delta R(Z,N)_{\mbox{def}}$ comes from the variance of deformation in the relevant nuclei,
\begin{eqnarray}
   \delta R(Z,N)_{\mbox{def}}  & \equiv& \frac{5}{8\pi}\delta(R_0\beta_2^2) \nonumber \\
   &=&  \sqrt{\frac{3}{5}}\frac{5}{8\pi}r_0[A^{1/3}\beta_2^2(Z,N)   \nonumber \\
   &&  + (A-4)^{1/3}\beta_2^2(Z-2,N-2)  \nonumber \\
   && -(A-2)^{1/3}\beta_2^2(Z-2,N)   \nonumber \\
   && - (A-2)^{1/3}\beta_2^2(Z,N-2)]   \nonumber \\ 
   &\approx&  \sqrt{\frac{3}{5}} \frac{5}{8\pi}r_0A^{1/3}\delta\beta_2^2 \;.
   \label{eq7}
\end{eqnarray}
The approximation of the last term is valid especially for heavier system. 
Because of the negligible contribution of $\delta R(Z,N)_{\mbox{sph}}$,  the resulting  $\delta R(Z,N)$ is mostly determined by the terms relevant to nuclear deformation. 
Although dynamic deformations and higher-order multipoles are not included in this equation,
they can be subsumed in principle under the deformation term in which $\delta\beta_2^2$ is replaced by $\sum_i\delta\beta_i^2$. 

\section{Correlation between charge radius and quadrupole deformation}

\subsection{Experimental data}
\begin{figure}[htbp]\noindent
\centering
\includegraphics[width=0.45\textwidth]{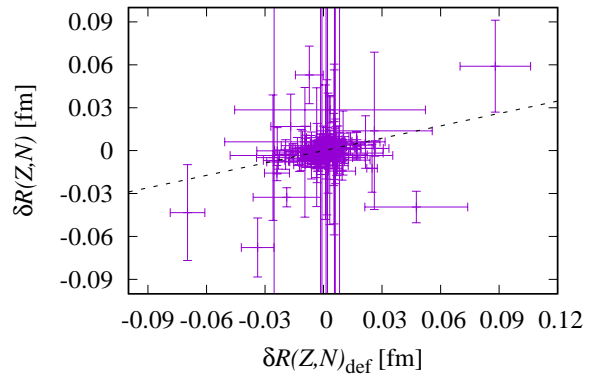}
 \caption{(Color online) $\delta R(Z,N)$ as a function of $\delta R(Z,N)(\beta_2^2)$ for all experimentally known cases. The linear fit is indicated by the dash line.}\label{fig.1}
\end{figure}
  
We can now examine the correlation between the experimental $\delta R(Z,N)$ and $\delta R(Z,N)_{\mbox{def}}$. The resulting correlation plot is shown in Fig.~\ref{fig.1}.
The experimental charge radius and deformation data are from the latest evaluations~\cite{Angeli2013,beta2016}.  There are in total 149 even-even nuclei from Ne to Cm. 
It is seen that almost all the data follow a linear trend within 1 standard deviation ($\sigma$).  A coefficiency of 0.29(6)  is determined with a reduced $\chi^2$ of 0.8. 
This indicates that experimental data of charge radii and quadrupole deformation are in well consistency.

\begin{figure}[htbp]\noindent
\centering
\includegraphics[width=0.45\textwidth]{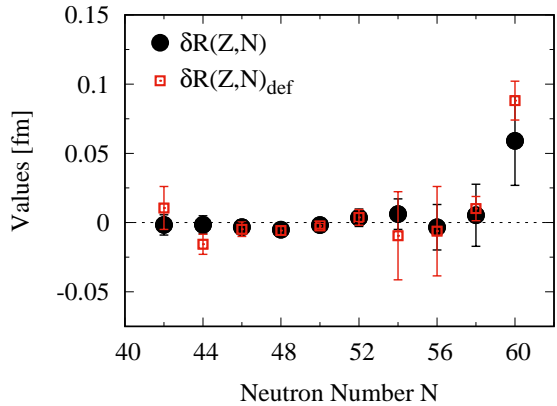}
 \caption{(Color online) Experimentally known $\delta R(Z,N)$ (filled circles) and  $\delta R(Z,N)_{\mbox{def}}$ (open squares) for the Sr isotopic chain.
 }\label{fig.2}
\end{figure}

Specifically, this correlation can be seen  for the Sr isotopes in Fig.~\ref{fig.2}. 
The sudden onset of the shape transition at $N=60$  is reflected distinctly by both $\delta R(Z,N)_{\mbox{def}}$ and $\delta R(Z,N)$. 
The deformation parameters for the relevant 4 nuclei, $^{98}_{38}$Sr, $^{96}_{38}$Sr, $^{96}_{36}$Kr and $^{94}_{36}$Kr are 0.40(1), 0.175(6), 0.25(3), 0.19(1),  resulting in 
 $\delta R(Z,N)_{\mbox{def}}$ of  0.088(14)  fm for $^{98}$Sr. 
Therefore, once considering the deformation correlation, the large $\delta R(Z,N)$ value at $N=60$, the well-known region of phase transitions, can be largely diminished.  
Similar correlation has been observed at $N\sim 90$ for the Nd isotopes. Unfortunately, it is not possible yet to test the region at $Z\sim80$ due to missing deformation data experimentally. 
 
5 cases at $^{20}_{10}$Ne, $^{46}_{20}$Ca, $^{46}_{22}$Ti,   $^{76}_{34}$Se, and  $^{78}_{36}$Kr,  show deviations from the linear trend by more than 2$\sigma$.  
Of them, $^{20}_{10}$Ne is the lightest nuclide with available charge radius and deformation data. It is known already 
that the precision of the charge radius formula deteriorates with decreasing mass number (in particularly for $A<60$)~\cite{Sun2014PhysRevC.90.054318}. 
This may be understood in the fact that the collective ``deformation'' property is more suitable for heavy nuclei in comparision with lighter nuclei. 
Further check on the mass dependence confirms this argument, where 8 of 10 cases  with $A<60$ have deviations from the linear trend by more than 1 $\sigma$. 
It should be noted that 1 $\sigma$ discussed here is even as small as 0.0040 fm. 

These few cases may be (partially) related to the various 
sources of charge radii and deformation, given the different methods of systematic errors associated with different techniques.
We noted that a recent analysis~\cite{BE2-Birch2016145} of $B(E2)$ measurements, from which $\beta_2$ is derived, concluded that most prevalent methods of measuring $B(E2)$ 
values are equivalent. 
Such comparison is not available yet for charge radius data across the entire chart of nuclides. 
Anyway, a consistent and equivalent set of nuclear charge radii and $B(E2)$ data are definitely crucial.  
A combined analysis of the cases for $^{46}_{20}$Ca and $^{46}_{22}$Ti shows that 
increasing the charge radius by 0.3\% or decreasing the $\beta_2$ by about 40\% (unlikely) for $^{44}_{20}$Ca 
 will result in their $\delta R(Z,N)$ values in better agreements with the linear trend.   
Similar arguments hold also for the cases of $^{76}_{34}$Se and $^{78}_{36}$Kr. 
Therefore, the correlation identified here may provide us a very accurate way
to investigate the consistence of both the charge radius and deformation surface. 

In Ref.~\cite{Sun2014PhysRevC.90.054318}, it was found that Eq.~(\ref{eq1}) is remarkably successful even at nuclei with magic neutron and/or proton numbers. 
This can be easily understood with the correlation identified. Nuclei with magic number of neutrons and/or protons is mostly in the spherical shape, i.e., 
their $\beta_2$ values of less than 0.1, thus leading to naturally a net  $\delta R(Z,N)$. 
This is very different from the counterpart in nuclear mass, the valence proton-neutron interactions $\delta V_{pn}$~\cite{Brenner2006PhysRevC.73.034315,Cakirli2005PhysRevLett.94.092501,Chen2009Phys.Rev.Lett.122503}. It depends strongly on the spatial overlap of the valence orbits and presents a dramatic variation when crossing neutron shell closures. 

\subsection{Correlations in nuclear models}
 
The knowledge of both experimental data on charge radii and deformation are still very limited. 
It will be very useful if nuclear models can provide these data either in absolute values or in differential values at a reasonable precision. For example, 
when one experimental deformation parameter is missing, one can resort to the theoretical predictions in nuclear models.
Care should be taken that quantities like $B(E2)$ refer to the charge (proton) distribution in the nucleus and that in particular $\beta$ is 
the charge deformation related to this charge distribution. This should be kept in mind when comparing the experimental results to nuclear models.

\begin{figure}[htbp!]\noindent
\centering
\includegraphics[width=0.45\textwidth]{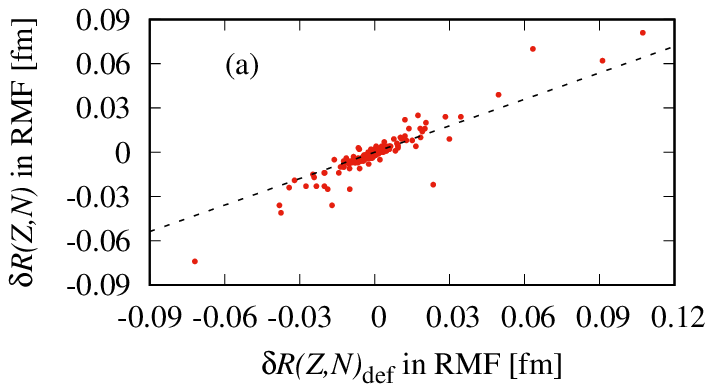} 
\includegraphics[width=0.45\textwidth]{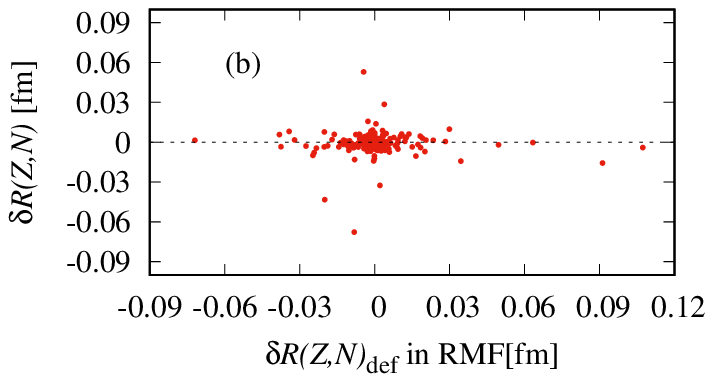}
 \caption{(Color online) Same as Fig.~\ref{fig.1} but using the predicted $\delta R(Z,N)$ and $\delta R(Z,N)_{\mbox{def}}$ in the RMF model (a), and 
 using experimental $\delta R(Z,N)$ and predicted $\delta R(Z,N)_{\mbox{def}}$ in the RMF model (b). Errorbars are not shown for experimental 
 data.}\label{fig.3}
\end{figure}

Global nuclear models can provide both charge radius and deformation data self-consistently. We choose Hartree-Fock-Bogoliubov (HFB-24) 
model~\cite{Goriely2013PhysRevC.88.024308} and the relativistic mean-field (RMF) model~\cite{Geng2005PTP} to check the same correlation.  As shown in Fig.~\ref{fig.3}(a), 
a linear correlation is predicted between $\delta R(Z,N)$ and $\delta R(Z,N)_{\mbox{def}}$ in the RMF.
The slope is determined to be 0.60, about a factor two larger than that of experimental data. 
The difference from the experimental trend should be related to the fact that all nuclei are treated with axially symmetric shapes in the RMF approach. 
The same correlation has also been found in the HFB-24 but with a different coefficience parameter (0.85).  

However, such correlation vanish once we use the theoretical $\beta_2$ values instead of experimental data. 
As an example, $\delta R(Z,N)$ vs. $\delta R(Z,N)_{\mbox{def}}$ calculated using the $\beta_2$ in the RMF is shown in Fig.~\ref{fig.3}(b). 
This indicates that current nuclear models are not accurate enough yet to reproduce the correlation seen in the experiment.

\section{Discussion}
\subsection{Improved charge radius formula}

As verified in the previous section, the $\delta R(Z,N)$ can be quantitatively reproduced with the $\delta R(Z,N)_{\mbox{def}}$ for the existing data. 
This leads to the improved charge radius formula as follows:
\begin{equation}\label{eq8}
\delta R(Z.N)^{\mbox{corr}}=\delta R(Z,N)-\mbox{C}\cdot \delta R(Z,N)_{\mbox{def}} \approx 0 \; 
\end{equation}
where $\mbox{C}$ is 0.29(6) determined from Fig.~\ref{fig.1}. $\delta R(Z,N)$ and $\delta R(Z,N)_{\mbox{def}}$ are given in Eq.~(\ref{eq2}) and  Eq.~(\ref{eq7}), respectively. 

The cases with experimental data are used to the check the accuracy of the relation without/with deformation correction. 
The weighted mean values of $\delta R(Z,N)^{\mbox{corr}}$ amount to only -8$\times 10^{-4}$ fm with the weighted standard deviation of 5$\times 10^{-3}$ fm. 
This is about 15\% improvement in precision comparing with that without correction (i.e., $\delta R(Z,N)$).  
The significance is that Eq.~(\ref{eq8}) can be extended to cases even when sudden variances occur in nuclear shapes.   
 
\subsection{Correlation of charge radius with $B(E2)$ and $\tau$}

Experimental quadrupole deformation values are derived from the model-independent 
experimental values of the reduced electric quadrupole transition probability $B(E2)$, between
the $0^+$ ground state and the first $2^+$ state in even-even nuclides, using the semi-empirical approach, 
\begin{equation}
\beta_2 = \frac{4\pi}{3ZR_0^2}[\frac{B(E2)}{\mbox{e}^2}]^{1/2} \;.
\label{eq:be2}
\end{equation}
Here $R_0=r_0A^{1/3}=1.2A^{1/3}$ fm and $B(E2)$ is in units of e$^2$b$^2$.
It is assuming a uniform charge distribution out to the distance $R$ and zero charge beyond~\cite{Greiner1996book,RAMAN2001ADNDT78}. 

The $B(E2)$ values are fundamentally important quantities for determining the collectivity in nuclei. 
Moreover, $B(E2)$ are related to  the  mean lifetime $\tau$ of the first $2^+$ state  through
\begin{equation}
  \tau(1+\alpha) = 40.81\times 10^{13}E^{-5}[\frac{B(E2)}{\mbox{e}^2\mbox{b}^2}]^{-1} \;,
\end{equation}
where $E$ is the excitation energy of the first $2^+$ state (in units of keV), and $\tau$ in ps. 
The total internal conversion coefficient $\alpha$ for a specific $E$ is needed for correction.  

$\delta R(Z,N)_{\mbox{def}}$ can be accordingly rewritten in terms of $B(E2)$ and $\tau$, 
\begin{eqnarray}
   \delta R(Z,N)_{\mbox{def}} &=&  4.35\times 10^3 \delta(\frac{B(E2)/ \mbox{e}^2\mbox{b}^2}{Z^2A})\mbox{b}^{1/2} \nonumber \\
   &=& 1.77\times 10^{21}\delta\left(\frac{E^{-5}Z^{-2}A^{-1}}{\tau(1+\alpha)}\right) {\mbox{b}^{1/2}} \;,
   \label{eq9}
\end{eqnarray}
where all the quantities $E$, $B(E2)$, $Z$, $A$ and $\alpha$ in the $\delta$ term are for the four neighboring even-even nuclei.
In case of no abrupt shape transition,   $\delta R(Z,N)_{\mbox{def}} \simeq 0$,  and the following relation involving again four neighboring doubly even nuclei should hold well, 
\begin{eqnarray}
 \delta(A^{1/3}\beta_2^2 ) \simeq \delta(\frac{B(E2)}{Z^2A}) \simeq  \delta\left(\frac{E^{-5}Z^{-2}A^{-1}}{\tau(1+\alpha)}\right) \simeq 0 \;.
\end{eqnarray}
For heavy nuclear system, where the difference in  ${Z^2A}$ can be safely neglected, we can then get the relation  
\begin{eqnarray}
   \delta B &= &B(E2)(Z,N)-B(E2)(Z,N-2)     \nonumber \\
                        & &-B(2)(Z-2,N)  +B(E2)(Z-2,N-2))    \nonumber \\
                        &\simeq & 0  \;,
   \label{eq11}
\end{eqnarray}
The same relation was proposed independently in Ref.~\cite{BE2relation-PhysRevC.12.2038}.   
To examine the validity of  eq.(\ref{eq11}), we use the same data set as in Fig.~1 were used to examine the validity of .

\begin{figure}[htbp]\noindent
\centering
\includegraphics[width=0.45\textwidth]{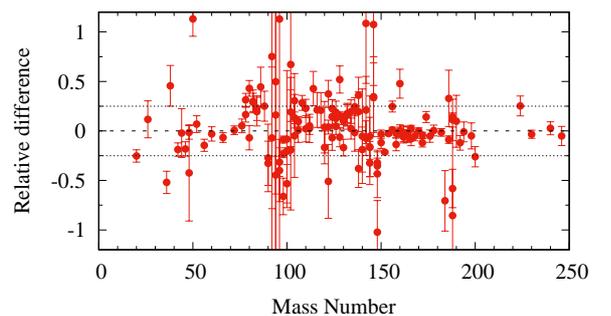}  
 \caption{(Color online)  Relative differences in $B(E2)$ between the predictions of Eq. (12) and experimental data. The $\pm$ 25\% accuracy is indicated by two dotted lines.}\label{fig.4}
\end{figure}

The ``theoretical'' $B(E2)$ value of a given nucleus $(Z,N)$,  $B(E2)_{\mbox{pred}}$,  is calculated  in
terms of experimental $B(E2)$ of its three neighboring nuclei $(Z-2,N)$, $(Z,N-2)$, $(Z-2,N-2)$.  Fig.~\ref{fig.4} shows the 
relative differences of the predictions defined as $[B(E2)_{\mbox{pred}}-B(E2)_{\mbox{exp}}]/B(E2)_{\mbox{exp}}$. 
It is seen that $B(E2)$  values can be calculated within an accuracy of $\pm$25\%, and often better. 
The large deviations are shown at $N\sim 60$ and 90. It should be aware that the precisions of experimental data are getting worse as well at these regions. 

Inserting Eq.~(\ref{eq9}) into Eq.~(\ref{eq8})  we get a correlation between four-point charge radius relation  with that of $B(E2)$ or $\tau$. 
This new relation, in principle,  should be more accurate for predictions of unknown  $B(E2)$ values than Eq.~(\ref{eq11}), because 
possible shape transitional effect can be at least partially compensated by the relevant charge radius data. 
When seven quantities out of  eight , e.g., four charge radii and three $B(E2)$ data, are known, then the last $B(E2)$ can be calculated. 
Unfortunately, this will give mostly a too large uncertainty to make meanful predictions. 
The uncertainty is mainly propagated from the charge radius data and is typically about 1 order of magnitude higher than that from Eq.~(\ref{eq11}).

\section{Conclusion}

With the available experimental data,  a linear correlation has been found between the charge radius relation $\delta R(Z,N)$ and 
the according quadrupole deformation (and thus $B(E2)$) relation $\delta R(Z,N)_{\mbox{def}}$. This correlation can provide a consistence check or analysis of experimental data on charge radii and deformation data.
In the near future it is also interesting to see whether the linear coefficiency 0.29(6) remains for more exotic nuclei, especially at shape transitional regions.  

The large deviation of four-point charge radius relation $\delta R(Z,N)$ at shape transitional regions, can be quantitatively reproduced 
with the $\delta R(Z,N)_{\mbox{def}}$ when experimental data are available. 
This in turn gives an improved charge radius formula, and therefore is very useful to make reliable short-range extrapolations of 
charge radii over the nuclear chart. Same correlation has been found in globle nuclear models, but so far the model itself is not accurate enough to reproduce the experimental data. Moreover, the relation can be generalized to a new relation between charge radius and $B(E2)$ or $\tau$. A simple four-point $B(E2)$ relation can reproduce experimental $B(E2)$ values within an accuracy of about $\pm$25\%. 
 
Finally, we would like to mention that a consistent description~\cite{Wood1999Nucl.Phys.A323,E0-PhysRevLett.101.022502,E0-PhysRevC.85.034331,PhysRevC.79.054301,PhysRevC.80.061301,PhysRevC.85.034321,Li2013866,Zhao2014Phys.Rev.C11301} 
of radius and transition probabilities of atomic nuclei are important to understand their correlation and thus for a better
interpretation of experimental results. A recent example is shown in $^{111-129}$Cd~\cite{Yordanov2016PhysRevLett.116.032501}, 
in which the parabolic behavior of the charge radii  is found due to the linear tendency of the quadruple deformation.

\section{Acknowledgments}
This work has been supported by the NSFC
under No. 11235002, 11475014 and National Program on Key Basic
Research Project (2016YFA0400502). The authors thank Z. P. Li, Z. M. Niu, P.W. Zhao, and L. H. Zhu for useful comments. 

%=================================================================
%\input{tbib.tex}
%\bibliography{../../ref}% Produces the bibliography via BibTeX.

%\end{CJK}
\end{document}